\documentclass[twocolumn,pra,superscriptaddress]{revtex4}
\usepackage{amsmath,amssymb,}
\usepackage{mathrsfs}
\usepackage{graphicx}
\usepackage{dcolumn}
\usepackage{bm}

\usepackage[colorlinks,citecolor=red]{hyperref}

\usepackage{rotating}
\usepackage{floatrow}

\usepackage{lipsum}

\usepackage{bbm}

\begin{document}

\title{Geometry and superfluidity of the flat band in a non-Hermitian optical lattice}

\author{Peng He}
\affiliation{National Laboratory of Solid State Microstructures
and School of Physics, Nanjing University, Nanjing 210093, China}
\author{Hai-Tao Ding}
\affiliation{National Laboratory of Solid State Microstructures
and School of Physics, Nanjing University, Nanjing 210093, China}

\author{Shi-Liang Zhu}
	\email{slzhu@nju.edu.cn}
\affiliation{Guangdong Provincial Key Laboratory of Quantum Engineering and Quantum Materials, School of Physics and Telecommunication Engineering, South China Normal University, Guangzhou 510006, China}

\affiliation{Guangdong-Hong Kong Joint Laboratory of Quantum Matter, Frontier Research Institute for Physics, South China Normal University, Guangzhou 510006,
China}

\date{\today}

\begin{abstract}
We propose an ultracold-atom setting where a fermionic superfluidity with attractive \textit{s}-wave interaction is uploaded in a non-Hermitian Lieb optical lattice. The existence of a real-energy flat band solution is revealed. We show that the interplay between the skin effect and flat-band localization leads to exotic localization properties. We develop a multiband mean-field description of this system and use both order parameters and superfluid weight to describe the phase transition. A relation between the superfluid weight and non-Hermitian quantum metric of the quantum states manifold is built. We find non-monotone criticality depending on the non-Hermiticity, and the non-reciprocity prominently enhances the phase coherence of the pairing field, suggesting ubiquitous critical behavior of the non-Hermitian fermionic superfluidity.
\end{abstract}

\maketitle

\section{Introduction}
A flat band is a dispersionless Bloch band with constant energy  and heavy degeneracy for all quasi-momentum. Typically, the flat band can arise as a result of destructive interference in bipartite quantum systems imposed by certain symmetries, or well designed coupling \cite{ZLiu2014,Leykam2018,Lieb1989,Vidal1998,Vidal2000}. Systems with flat bands host intriguing features which are entirely governed by the quantum geometry and topology of the bands \cite{KSun2011,Peotta2015,Rhim2019,Julku2016,Torma2018,Raoux2014,Piechon2016,WJiang2019,Rhim2020,Danieli2020,DSMa2020,Kuno2020,
CSChiu2020}, facilitating the correlation in superconductivity \cite{ETang2014,Mondaini2018,JMao2020,Swain2020}, fractional quantum Hall states \cite{Neupert2011,Bregholtz2013}, and frustration in ferromagnetism \cite{Lieb1989,Mielke1991}. For instance, the flat band yields a maximal critical temperature within the mean-field theory of superconductivity \cite{Kopnin2011,Iglovikov2014}, and plays an important role in the cuprates \cite{Emery1987} and twisted-bilayer graphene \cite{FXie2020,XHu2019,Julku2020}. The flat bands have been recognized and experimentally observed in  various contexts of condensed matter and synthetic quantum matter. Specifically, the flat bands in Lieb lattice and kagome lattice have been realized with the ultracold gases \cite{Taie2015,Leung2020,GWChern2014,Jo2012}, photonic crystals \cite{Silva2014,Mukherjee2015,Vicencio2015}, and electronic systems \cite{Slot2017}.  However, it is still an interesting direction to test flat-band physics in many varieties and classifications of states and phases showing progress of experiments.

In recent years, non-Hermitian states of matter have attracted considerable attention both theoretically and experimentally \cite{Ghatak2019,Bergholtz2019,Ashida2020}. The systems described by non-Hermitian Hamiltonians are usually non-conserved, such as solids with finite quasi-particle lifetimes \cite{Kozii2017,HShen2018,Papaj2019,XShen2019}, artificial lattice \cite{DWZhang2018} with gain and loss or nonreciprocity \cite{JLi2019,HZhou2018,Cerjan2019,DWZhang2020,HJiang2019,LZTang2020,LJLang2021,LJLang2018}, etc. Recent developments have  revived interest in various physical aspects. The flat band has also been proposed in non-Hermitian systems \cite{SMZhang2019,LJin2019,Leykam2017,LGe2018,Ramezani2017,Maimaiti2020}, but still remains largely unexplored.

In this work, we study fermionic superfluidity with attractive \textit{s}-wave interaction in an optical Lieb lattice. With the inclusion of both atom loss and inelastic collision, together with an auxiliary lattice, the system is effectively described by a non-Hermitian Hamiltonian with non-reciprocal hopping amplitudes and complex interaction strength. The Lieb lattice features a diabolic single Dirac cone intersected with a flat band \cite{RShen2010,Weeks2010}. In the presence of non-Hermiticity, the Dirac point will extend into a pair of exceptional points (EPs). As suggested by previous works, the non-Hermitian system with nonreciprocal hopping amplitudes will exhibit the skin effect of which all bulk bands are pumped at the boundaries, as a manifestation of point gap topology associating with the exceptional points \cite{SYao2018,Borgnia2020,FSong2019,KZhang2020,LLi2020,Okuma2020,Jin2019}. We propose a non-Hermitian flat band localization coexisting with the skin effect. We demonstrate that the skin effect is forbidden by the geometric frustration of particle motion for the flat band. Interestingly, the non-reciprocity enhances the flat band localization. Distinguishing from the skin effect, the enhancement does not rely on the boundary condition. This implies nontrivial physics enforced with the nonreciprocity even under periodic boundary condition which does not support the skin effect. Furthermore, we find that the nonreciprocity prominently enhances the phase stiffness of the fermionic superfluidity. We adopt mean-field description of the fermionic superfluidity by generalizing the non-Hermitian  Bardeen-Cooper-Schrieffer (BCS) theory developed in Ref. \cite{Yamamoto2019} to the multiband case, leading to a non-Hermitian Bogoliubov–de Gennes (BdG) Hamiltonian. In general, the emergence of the superconducting phase does not only require the form of the pairing field, but also the phase coherence of the pairs, which enables the Meissner effect \cite{Corson1999}. We solve the gap equation and calculate the superfluid weight. We show that the superfluid weight for the non-Hermitian superfluidity is related with the integral of non-Hermitian metric tensor of the quantum state manifold over the Brillouin zone, which is a manifestation of nontrivial flat-band effect.

This paper is organized as follows. In Sec. \ref{sec_model}, we propose an ultracold-atom-based setup and present an effective non-Hermitian Hamiltonian. In Sec. \ref{sec_free}, the band structures and the localization properties of the states are addressed in terms of interplay between the skin effect and destructive interference. Section \ref{sec_BCS} develops a multiband mean-field description of the fermionic superfluid, and solves the gap equation. In Sec. \ref{sec_sw}, we compute the superfluid weight and elucidate the unique criticality of the phase transition. Finally, a short conclusion is given in Sec. \ref{sec_conclusion}.

\begin{figure}[htbp]
	\centering
	\includegraphics[width=\textwidth]{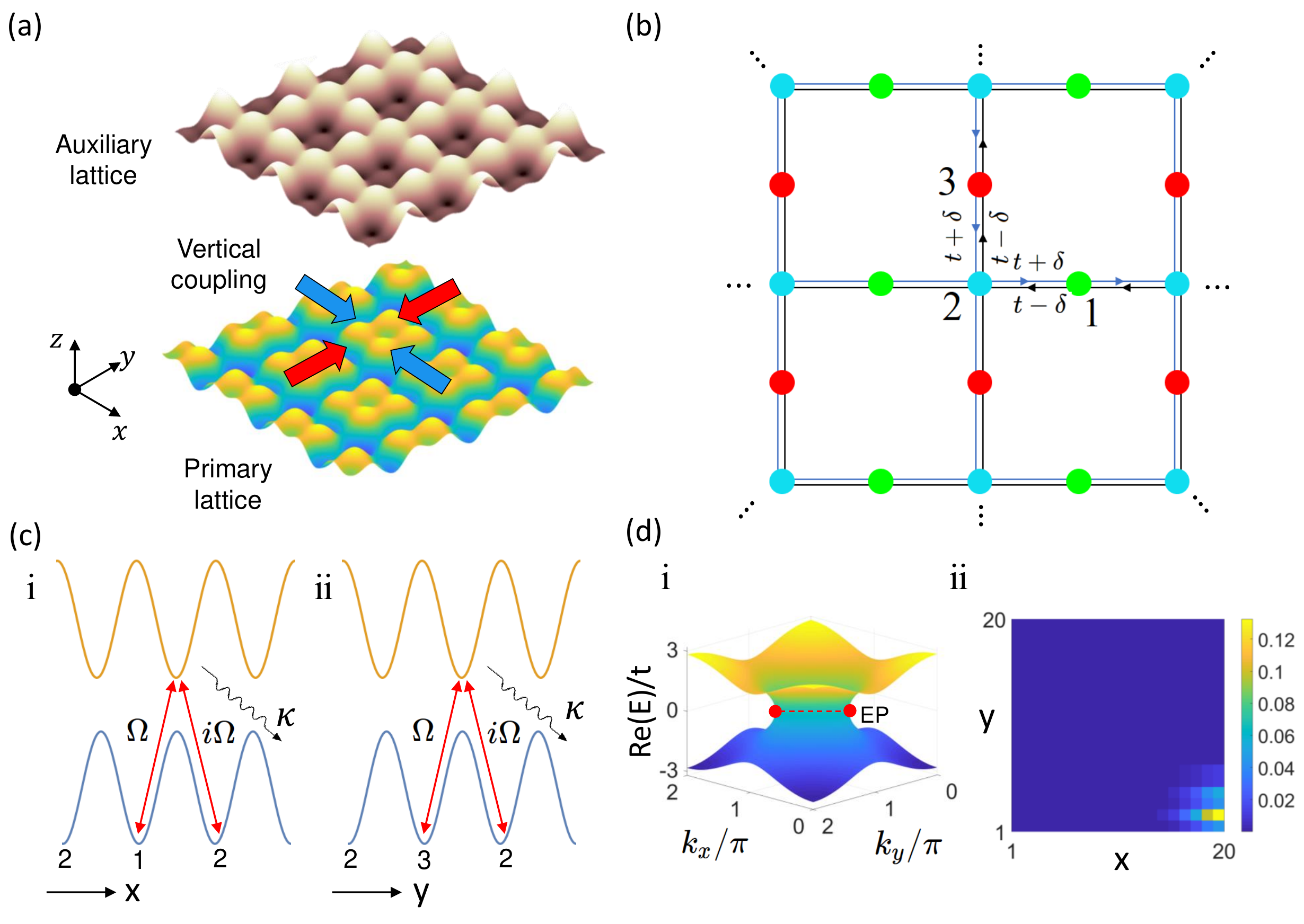}
	\caption{(a) Schematic illustration of our proposed experimental setup with ultracold atoms in an optical lattice. (b) A Lieb lattice with nonreciprocal hopping, which has three inequivalent sites in one unit cell.  (c) The cross sections of (a) along (i) the x direction and (ii) the y direction, with sublattices 1,2 and 2,3 respectively. $\kappa$ is the on-site decay rate in the auxiliary lattice. (d), (i) The real part of the energy spectrum for $\delta/t=0.5$. Here we only show the two dispersive bands.  Two exceptional points are connected by a bulk Fermi arc (highlighted by the red dashed line). (ii) The real space density distribution of an eigenstate for the dispersive bands under the open boundary condition along the x and y directions.}
	\label{fig1}
\end{figure}

\section{Model Hamiltonian}\label{sec_model}

We consider a gas of fermionic atoms in a two-dimensional optical lattice $V_{\mathrm{prm.}}$ with coherent coupling to the auxiliary degrees of freedom, as illustrated in Fig. \ref{fig1}. The primary lattice has a line-centered geometry, which could be experimentally realized by superimposing three pairs of laser beams with the following formation \cite{Taie2015,RShen2010,Weeks2010},
\begin{equation}
\begin{split}&V_{\mathrm{prm.}}(x, y)=-V_{\mathrm{long}}^{(x)} \cos ^{2}\left(k_{\mathrm{L}} x\right)-V_{\mathrm{long}}^{(y)}\cos ^{2}\left(k_{\mathrm{L}} y\right)\\& -V_{\mathrm{short}}^{(x)} \cos ^{2}\left(2 k_{\mathrm{L}} x\right)-V_{\mathrm{short}}^{(y)}\cos ^{2}\left(2 k_{\mathrm{L}} y\right) \\& -{V_{\mathrm{diag}}^{(x)} \cos^{2}\left(k_{\mathrm{L}}(x-y)\right)}- V_{\mathrm{diag}}^{(y)}\cos ^{2}\left(k_{\mathrm{L}}(x+y)\right)\,,\end{split}
\label{eq_pot}\vspace{0.1 in}
\end{equation}
where $k_L=2\pi/\lambda$ is the wave number of the lattice and $\lambda$ the wavelength of the lasers. The potential amplitudes $V_{\mathrm{long}}^{(x,y)}$, $V_{\mathrm{short}}^{(x,y)}$ and $V_{\mathrm{diag}}^{(x,y)}$ could be tuned by adjusting the laser intensities, $\phi_{x,y}$ and $\varphi$ are the phases of laser beams. For simplicity we choose $(V_{long}^{(x)},V_{short}^{(x)},V_{diag}^{(x)})=
(V_{long}^{(y)},V_{short}^{(y)},V_{diag}^{(y)})=(V_{long},V_{short},V_{diag})$. To ensure the nearest-neighbor sites are coupled, the potential strength of the auxiliary lattice is tuned opposite to that of the primary lattice. The auxiliary lattice with decay is included to induce asymmetric hopping between the coupled sites, which generates the skin effect as we will see below.

Two equally populated magnetic sublevels of the hyperfine ground-state manifold are uploaded, to mimic the interacting spin-up and spin-down electrons moving in the lattice. Following Refs. \cite{ZGong2018,TLiu2019,Muller2012,Diehl2011}, we propose to engineer a collective loss  from a combination of on-site one-body losses of auxiliary lattice and two-body dissipations of the primary states. In the experiment, the one-body loss could be generated by applying a radio frequency pulse to resonantly transfer the atoms to an irrelevant excited state, while the two-body dissipation could be induced by inelastic collisions with photoassociation process \cite{Tomita2017}.

The full open-system dynamics in the rotating frame of reference can be written as
\vspace{0.1 in}
\begin{widetext}
\begin{equation}
\begin{split}
\dot{\rho}_{t}=&-i [H_{0}+\frac{\Omega}{2} (\sum_{i,j,\sigma} a_{i, j,2,\sigma}^\dagger (c_{i, j, 1,\sigma}+i c_{i, j, 2,\sigma} )+a_{i, j, 1,\sigma}^\dagger (c_{i, j, 1,\sigma}+i c_{i+1, j, 2,\sigma} )
 +a_{i, j, 3,\sigma}^\dagger (c_{i, j, 2,\sigma}+i c_{i, j, 3,\sigma} )\\ &+a_{i, j, 2,\sigma}^\dagger (c_{i, j, 2,\sigma}+i c_{i+1, j, 3,\sigma} )+\mathrm{H.c.} ), \rho_{\mathrm{t}} ]+ \sum_{\mathrm{i}, \mathrm{j}, \alpha} \mathcal{D} [L(a_{\mathrm{i}, \mathrm{j}, \alpha,\sigma},c_{\mathrm{i}, \mathrm{j}, \alpha,\sigma}) ] \rho_{\mathrm{t}}\,
\end{split}
\end{equation}
\end{widetext}
where $a_{i,j,\alpha,\sigma}$'s ($a_{i,j,\alpha,\sigma}^\dagger$'s) denote the annihilation (creation) operators of the particles with spin $\sigma=\uparrow,\downarrow$ in the auxiliary orbital centered at $\mathbf{r}_{i,j,\alpha}$ ($\alpha$ is the sublattice label), $c_{i,j,\alpha,\sigma}$'s ($c_{i,j,\alpha,\sigma}^\dagger$'s) denote corresponding annihilation (creation) operators of primary degrees of freedom,  $\Omega$ is the coupling strength between the primary lattice and auxiliary lattice, $H_0$ is the Hamiltonian for the primary lattice, and $\mathcal{D}[L]\rho\equiv L\rho L^\dagger -\frac{1}{2}\{L^\dagger L,\rho\}$ is the Lindblad superoperator with the $L$ relating to the operators $a_{i,j,\alpha,\sigma}$ and $c_{i,j,\alpha,\sigma}$.

In the regime that the on-site decay rate $\kappa\gg \Omega$, one can adiabatically eliminate the decay modes in the auxiliary lattice \cite{TLiu2019}. Thus the effective dynamics is well described by
\begin{equation}
\dot{\rho}_t=-i[\tilde{H}_0,\rho_t]+\tilde{\mathcal{D}}[L]\rho_t\,,
\end{equation}
\begin{equation}
\tilde{H}_0=-\sum_{\langle\mathbf{i}\alpha,\mathbf{j}\beta\rangle,\sigma}t_{\mathbf{i}\alpha,\mathbf{j}\beta}^\sigma c_{\mathbf{i}\alpha\sigma}c^\dagger_{\mathbf{j}\beta\sigma}-\tilde{U}\sum_{\mathbf{i}\alpha}n_{\mathbf{i}\alpha\uparrow}n_{\mathbf{i}\alpha\downarrow}\,,
\end{equation}
\begin{equation}
L=\sum_{<\mathbf{i}\alpha,\mathbf{j}\beta>,\sigma}\sqrt{2\gamma}(c_{\mathbf{i}\alpha\sigma}+ic_{\mathbf{j}\beta\sigma})+\sum_{\mathbf{i}\alpha}\sqrt{\Gamma}c_{\mathbf{i}\alpha\downarrow}c_{\mathbf{i}\alpha\uparrow}\,,
\end{equation}
where $\gamma=\Omega^2/(2\kappa)$ and $\Gamma$ is the two-body loss rate in the primary lattice. We only consider the tight-binding regime thus $\langle \mathbf{i}\alpha,\mathbf{j}\beta\rangle$ runs over all nearest-neighbor sites [see Fig. \ref{fig1} (b)], and we redefine $\mathbf{i}=(i_x,i_y)=(i,j)$.

If we only consider the dynamics over a short time, the quantum jump term $\gamma L\rho_t L^\dagger$ is negligible. Thus we have an effective non-Hermitian Hamiltonian $H_{\mathrm{eff}}=H_0-(i/2)L^\dagger L$ \cite{Yamamoto2019,Poyatos1996}. With Fourier transformation $c_{\mathbf{k}\alpha\sigma}=1/\sqrt{N_c}\sum_{\mathbf{i}} e^{-i\mathbf{k}\cdot\mathbf{r}_{\mathbf{i}\alpha}} c_{\mathbf{i}\alpha\sigma}$, the Hamiltonian in momentum space reads $\mathcal{H}_{\mathrm{eff}}=\mathcal{H}_{\mathrm{kin}}+\mathcal{H}_{\mathrm{int}}-\mu N$. The kinetic term $\mathcal{H}_{\mathrm{kin}}$ is given by $\mathcal{H}_{\mathrm{kin}}=\sum_{\mathbf{k}} \Psi_{\mathbf{k}}^\dagger H_\mathbf{k} \Psi_{\mathbf{k}}$, with $\Psi_{\mathbf{k}}=[c_{\mathbf{k}\alpha\uparrow},c_{\mathbf{k}\alpha\downarrow}]^\mathrm{T}$,
and
\begin{equation}
H_\mathbf{k\uparrow}=H_\mathbf{k\downarrow}=a_\mathbf{k}\lambda_1+b_\mathbf{k}\lambda_6\,,\label{eq_lieb}
\end{equation}
where $a_\mathbf{k}=2t\cos(k_x/2)-2i\delta\sin(k_x/2)$, $b_\mathbf{k}=2t\cos(k_y/2)+2i\delta\sin(k_y/2)$ ($\delta\equiv\gamma$), and $\lambda_{i}$ is the i-th Gell-Mann matrix. The interaction term is given by $\mathcal{H}_{\mathrm{int}}=-U\sum_{\mathbf{k}\mathbf{k'},\alpha} c^\dagger_{\mathbf{k}\alpha\uparrow}c^\dagger_{\mathbf{-k}\alpha\downarrow}c_{\mathbf{-k'}\alpha\uparrow}c_{\mathbf{k'}\alpha\downarrow}$, where interaction strength $U\equiv \tilde{U}+i\Gamma/2$ becomes complex-valued due to the inelastic scattering. We note that our model can be regarded as a non-Hermitian extension of the Hubbard model on a Lieb lattice, which is proposed and studied in Ref. \cite{Julku2016}.

\section{Flat band localization and skin effect}\label{sec_free}

\begin{figure}[htbp]
	\centering
	\includegraphics[width=\textwidth]{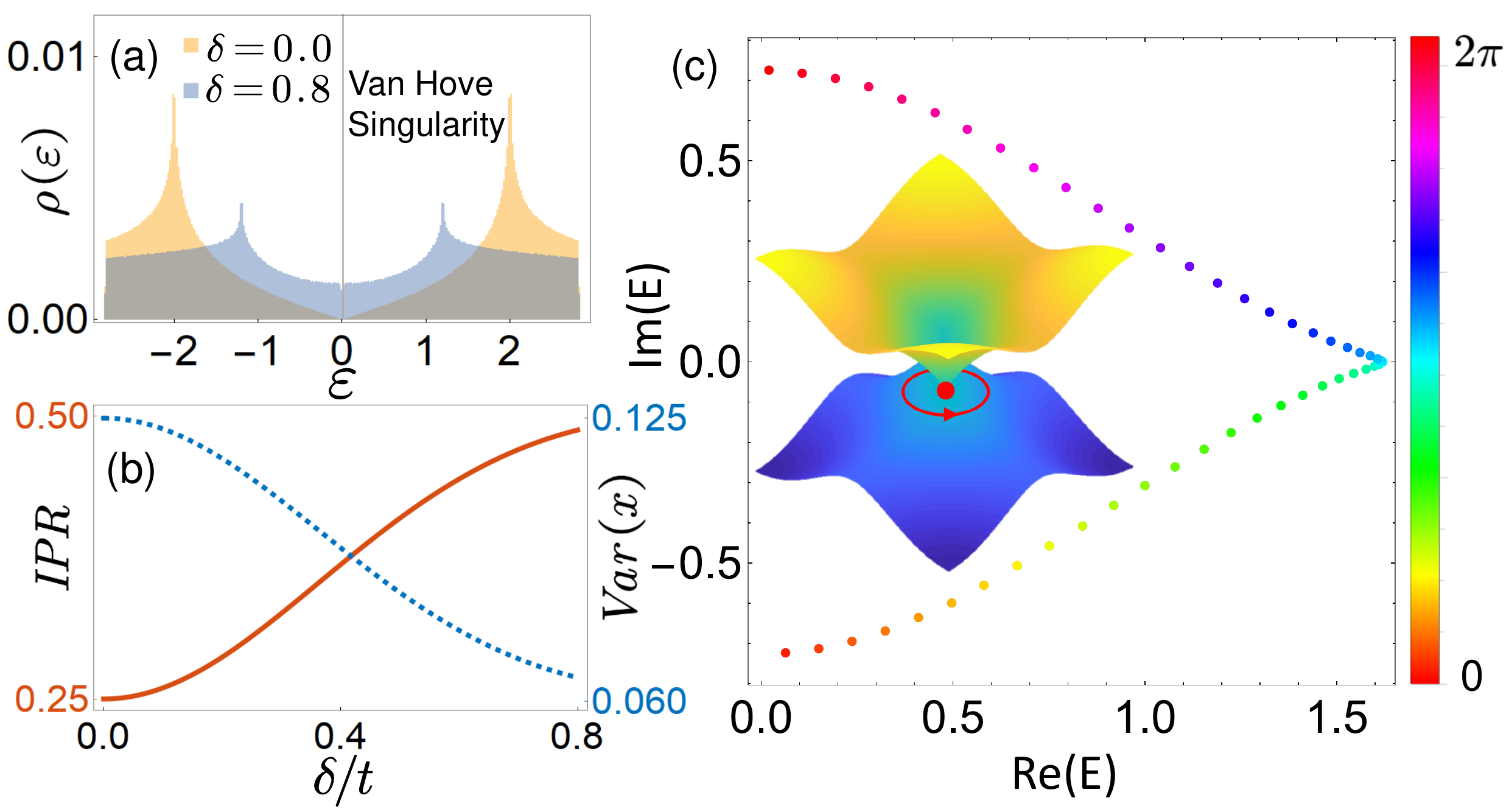}
	\caption{ (a) The density of states $\rho(\varepsilon)$ of the model described by Eq. (\ref{eq_lieb}) for  nonreciprocity $\delta=0$ and $\delta=0.8$, respectively. (b) The IPR (solid line) and the position variance (dashed line) for the ring-mode state. Here we take lattice constant $a=1$. (c) The swapping of energy eigenvalues along the loop $\mathcal{L}$ encircling the EP. The red line in the inset shows a path from $\theta=0$ to $\theta=2\pi$. $\theta \in [0,2\pi]$ parametrizes the loop $\mathcal{L}$. }
	\label{fig2}
\end{figure}

Let us start by considering the interaction-free Hamiltonian $H_\mathbf{k}$. The eigenequation is given by $H_\mathbf{k}|g_{n\mathbf{k}}\rangle=\varepsilon_{n\mathbf{k}}|g_{n\mathbf{k}}\rangle$ with the Bloch functions $|g_{n\mathbf{k}}\rangle$ and the energy dispersion $\varepsilon_{n\mathbf{k}}$ ($n=\pm,0$). The upper and lower bands have complex-valued dispersions $\varepsilon_{\pm}=\pm\sqrt{a_{\mathbf{k}}^2+b_{\mathbf{k}}^2}$ [as illustrated in Fig. \ref{fig1} (d,i)], while the middle band is purely real and strictly flat $\varepsilon_0=0$, as in the Hermitian limit. This is a result of the chiral symmetry of the Hamiltonian preserved under the non-Hermitian perturbation $\hat C H_\mathbf{k} \hat C=-H_\mathbf{k}$, with $\hat C\equiv\mathrm{diag}[1,-1,1]$. In contrast, for Lieb lattice considered in Ref. \cite{SMZhang2019}, which is in the presence of particle gain and loss on the sublattice sites, the non-Hermitian perturbation $H_{NH}=-i\gamma\mathrm{diag}[1,0,-1]$ breaks the sublattice symmetry. However, the flat band still could be restored from this system by virtue of artificial gauge potential and next-nearest-neighbor hopping.

The Lieb lattice features a single Dirac point at the corner of the Brillouin zone. In the presence of non-Hermiticity, the Hamiltonian $H_\mathbf{k}$ has a pair of EPs residing at $\mathbf{K}_{\pm}=\pm(\arccos(\frac{\delta^2-t^2}{\delta^2+t^2}),\arccos(\frac{\delta^2-t^2}{\delta^2+t^2}))$, which are connected by a bulk Fermi arc, as shown in Fig. \ref{fig1} (d,i). We calculate the density of states $\rho(\varepsilon)=\frac{1}{(2\pi)^2}\int d\mathbf{k} \delta(\varepsilon-\varepsilon_{\mathbf{k}})$ in Fig. \ref{fig2} (a). The density of the states near the Fermi surface (here we set $\mu=0$) increases for nonzero non-reciprocity $\delta$, which is expected due to the existence of the bulk Fermi arc.

As a bipartite quantum system from the chiral universality classes, the Lieb lattice admits destructive interference of states on the flat band, which guarantees coherent cancellation of net particle flows thus leads to the localization of states \cite{ZLiu2014,ZHZhang2016}. A single-particle state $|\Phi\rangle=\sum_{\mathrm{r},\alpha} \mathcal{P}_{\mathbf{r},\alpha} c_{\mathbf{r},\alpha}^\dagger|0\rangle$ satisfies $H_L|\Phi\rangle=0$ if and only if
\begin{equation}
\sum_{\langle(\mathbf{r},\alpha),(\mathbf{r}',\beta)\rangle} \mathcal{P}_{\mathbf{r}',\beta}=0\,,\quad \forall \quad \mathbf{r},\alpha\,,\label{eq_dstint}
\end{equation}
where $H_L$ is the lattice Hamiltonian in real space with $U=0$ and $|0\rangle$ is the vacuum state. We call the state satisfying condition Eq.(\ref{eq_dstint}) a ring mode state. To see this, we consider straightforward manipulations on Wannier states,
\begin{equation}
H_L c_{\mathbf{i}1}^\dagger|0\rangle=[(t-\delta)c_{\mathbf{i}-\frac{a}{2}i_x,2}^\dagger+(t+\delta)c_{\mathbf{i}+\frac{a}{2}i_x,2}^\dagger]|0\rangle\,,
\end{equation}
\begin{equation}
H_L c_{\mathbf{i}3}^\dagger|0\rangle=[(t+\delta)c_{\mathbf{i}-\frac{a}{2}i_y,2}^\dagger+(t-\delta)c_{\mathbf{i}+\frac{a}{2}i_y,1}^\dagger]|0\rangle\,,
\end{equation}
thus we have a ring-mode state with $\varepsilon=0$,
\begin{equation}
	 |\Phi\rangle=\mathcal{N}[t_1c_{i,j,3}^\dagger-t_2c_{i,j,1}^\dagger
+t_2c_{i+1,j+1,3}^\dagger-t_1c_{i+1,j+1,1}^\dagger]|0\rangle\,,\label{eq_RM}
\end{equation}
where $t_1=\sqrt{\frac{t-\delta}{t+\delta}}$, $t_2=\sqrt{\frac{t+\delta}{t-\delta}}$, and $\mathcal{N}$ is the normalization factor. Furthermore, any linear superposition of the single-plaquette ring-mode states given by Eq.(\ref{eq_RM}) is also a solution, which gives arise to a large subspace with $\varepsilon=0$. As revealed in Ref. \cite{Mielke1991}, this can also be understood by the line graph theory. A line graph $L(G)$ of a graph $G$ is constructed with its vertex set being the edge set of $G(V,E)$. The incidence matrix $B(G)=(b_{ve})_{|V|\times|E|}$ (with unit entry $b_{ve}=1$ if the vertex $v$ is incident to the edge $e$ and zero otherwise) is related with the adjacency matrix $A_L$ of the line graph by $B^TB=2I+A_L$, due to the fact that every edge is always connected with two vertexes. If the number of vertexes $n_V$ is smaller than that of the edges $zn_V/2$, $A_L$ will have a highly degenerate subspace with eigenvalue $-2$, such that $B(G)$ has large zero space.

To characterize the flat band localization, we evaluate the inverse participation ratio (IPR) $I_{f.b.}$ and the variance $\mathrm{Var}(\hat{\mathbf{r}})$ of the position operator $\hat{\mathbf{r}}$ \cite{Kuno2020,DWZhang2020,Ozawa2019}, and the results are shown in Fig. \ref{fig2} (b). The real space IPR is defined as $I_{f.b.}=\sum_{\mathbf{i},\nu}|\Psi_{\mathbf{i},\nu}|^4$, where $|\Psi_{\mathbf{i},\nu}|^2$ is the local density of the state on single site $\mathbf{r}_{\mathbf{i},\nu}$. For the ring-mode state $|\Phi\rangle$ given by Eq. (\ref{eq_RM}), $I_{f.b.}= \frac{\delta^4+6 \delta^2 t^2+t^4}{4 (\delta^2+t^2)^2}$. The IPR exhibits power-law increase for small nonreciprocity, $I_{f.b.} \sim \frac{1}{4}+\frac{\delta^2}{t^2}+\mathcal{O}(\delta^3 )$, which implies that the non-reciprocity enhances the flat band localization. The localization property can also be characterized by the position variance $\mathrm{Var}(\hat{\mathbf{r}})=\langle \Phi|\hat{\mathbf{r}}^2|\Phi\rangle-\langle \Phi|\hat{\mathbf{r}}|\Phi\rangle^2$. It can be readily obtained that $\mathrm{Var}(\hat{x})=\mathrm{Var}(\hat{y})=\frac{a^2 (\delta^4+t^4)}{8 (\delta^2+t^2)^2}$ for the ring-mode state $|\Phi\rangle$. We illustrate $\mathrm{Var}(\hat{x})$ in Fig. \ref{fig2} (b). $\mathrm{Var}(\hat{x})$ decreases for the increasing $\delta$, confirming the same physics from the IPR.  We also note that the enhancement of localization does not rely on the boundary condition. This manifests nontrivial physics induced by the nonreciprocity even under periodic boundary condition, while the skin effect only occurs for open systems.

Due to the asymmetric hopping amplitudes, bulk states of non-Hermitian systems exhibit  anomalous boundary localization, dubbed the skin effect. As shown in Fig. \ref{fig1} (d,ii), all states with nonzero energy are pumped towards the direction with stronger hopping amplitude, which gives rise to corner-state distributions. This is reminiscent of the second-order skin effect studied in Ref. \cite{CHLee2019}, but only happens for the two dispersive bands. The skin effect in our model occurs as manifestation of the nontrivial point-gap topology signaling with the presence of the EPs, which is assigned with a winding number
\begin{equation}
\nu=\frac{1}{2\pi}\oint_\mathcal{L}\nabla_{\mathbf{q}} \arg[\varepsilon_+-\varepsilon_-]\cdot d\mathbf{q}\,,
\end{equation}
where $\mathcal{L}$ is a closed loop encircling the EP. Each EP is associated with $\nu=\pm\frac{1}{2}$, as illustrated in Fig. \ref{fig2}. The bands with a point gap (finite interior in the complex spectra) have nonzero direct current $J=\oint n(E,E^*) dE$ with $n(E,E^*)$ the energy distribution, which gives rise to nonreciprocal pumping, \emph{i.e.}, the skin effect \cite{KZhang2020}. In contrast, The flat band is purely real, and does not acquire a point gap in the complex plane, thus does not have the skin effect.

\section{Mean-field description}\label{sec_BCS}

\begin{figure}[htbp]
	\centering
	\includegraphics[width=\textwidth]{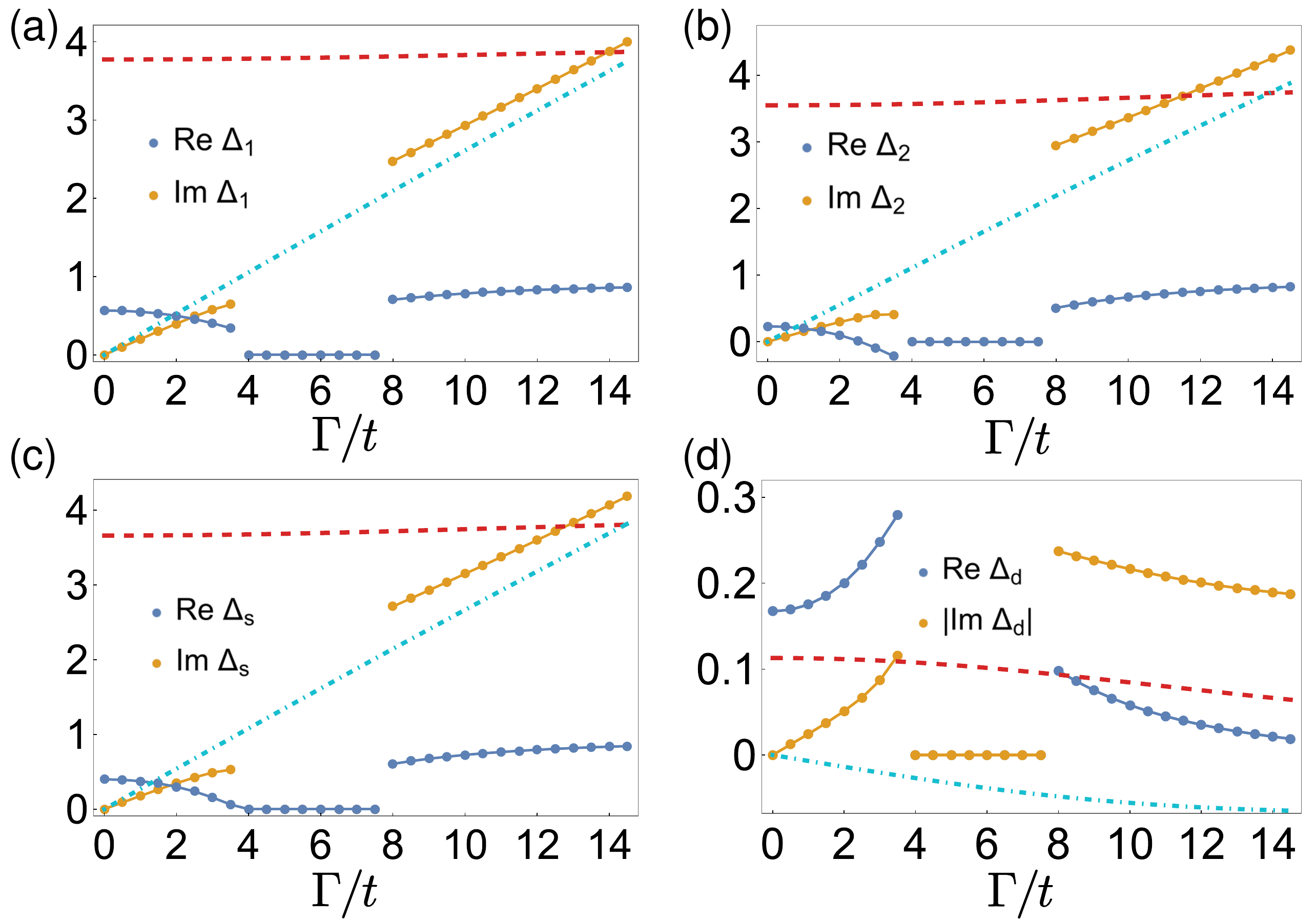}
	\caption{ Numerical solution of the order parameters as the function of $\Gamma$ for $\tilde{U}=1.8$ at zero temperature. The dashed red lines indicate the real part of the order parameters at strong interaction strength $\tilde{U}=10$, while the dotted dashed blue lines indicate the imaginary part.  }
	\label{fig3}
\end{figure}

We adopt a mean-field description within the framework of BCS theory and decouple the interaction term as
\begin{equation}
\mathcal{H}_{\mathrm{int}}\approx \sum_{\mathbf{k},\alpha} \Delta_{\alpha}c^\dagger_{\mathbf{k}\alpha\uparrow}c^\dagger_{\mathbf{-k}\alpha\downarrow}+
\bar{\Delta}_{\alpha}c_{\mathbf{-k}\alpha\uparrow}c_{\mathbf{k}\alpha\downarrow}\,,
\end{equation}
where $\Delta_{\alpha}=-U\sum_\mathbf{k}\langle\langle c_{\mathbf{-k}\alpha\downarrow}c_{\mathbf{k}\alpha\uparrow}\rangle\rangle$ and $\bar{\Delta}_{\alpha}=-U\sum_\mathbf{k}\langle\langle c_{\mathbf{k}\alpha\uparrow}^\dagger c_{\mathbf{-k}\alpha\downarrow}^\dagger\rangle\rangle$, and here we use the inner product in the biorthonormal basis $\langle\langle \mathcal{O} \rangle\rangle=\langle\tilde{\Psi}|\mathcal{O}|\Psi\rangle$, with $|\Psi\rangle$ ($\langle\tilde{\Psi}|$) the right (left) ground eigenstate of the Hamiltonian $\mathcal{H}_{\mathrm{eff}}$ and $\mathcal{O}$ any operator. In Nambu representation $\Psi_{\mathbf{k}}=[c_{\mathbf{k}\alpha\uparrow},c_{\mathbf{-k}\alpha\downarrow}^\dagger]^\mathrm{T}$, the mean-field Hamiltonian reads $\mathcal{H}_{\mathrm{MF}}=\sum_{\mathbf{k}}\Psi_{\mathbf{k}}^\dagger \mathcal{H}(\mathbf{k})\Psi_{\mathbf{k}}$, with the BdG Hamiltonian
\begin{equation}
\mathcal{H}(\mathbf{k})=\left[\begin{array}{cc}
H_{\mathbf{k}\uparrow}-\mu\mathbbm{1} & \Delta\\
\bar{\Delta}  & -(H_{\mathbf{k}\downarrow}+\mu\mathbbm{1})
\end{array}\right]\,,
\end{equation}
where $\mathbbm{1}$ is the $3\times3$ identity matrix, $\Delta=\rm{diag}[\Delta_1,\Delta_2,\Delta_3]$. Consistent with the spontaneous U(1) symmetry breaking of the BCS ground states, $\Delta$ could see a U(1) phase difference from $\bar{\Delta}$: $\Delta=\Delta_0 e^{i\theta}$ and $\bar{\Delta}=\Delta_0 e^{-i\theta}$. Henceforth we choose the gauge for which $\mathcal{H}_{\mathrm{MF}}^\dagger=\mathcal{H}_{\mathrm{MF}}^*$ to eliminate $\theta$, such that $\Delta=\bar{\Delta}$, as discussed in Ref. \cite{Yamamoto2019}. 

To obtain the superfluid weight, we consider the system subject to a uniform gauge field $\mathbf{A}$. By Peierls substitution $\mathbf{k}$ to $\mathbf{k}-\mathbf{q}$ ($\mathbf{q}=e\mathbf{A}$) and diagonalizing the kinetic block, the Hamiltonian reads
\begin{equation}
\bar{H}_{\mathbf{k}}(\mathbf{q})=\left[\begin{array}{cc}
\varepsilon_{\mathbf{k}-\mathbf{q}}-\mu \mathbbm{1} & \tilde{\mathcal{G}}_{\mathbf{k}-\mathbf{q}} \Delta \mathcal{G}_{\mathbf{k}+\mathbf{q}} \\
\tilde{\mathcal{G}}_{\mathbf{k}+\mathbf{q}} \Delta \mathcal{G}_{\mathbf{k}-\mathbf{q}} & -\left(\varepsilon_{\mathbf{k}+\mathbf{q}}-\mu \mathbbm{1}\right)
\end{array}\right]\,,\label{eq_BdG}
\end{equation}
where $\varepsilon_{\mathbf{k}}=\rm{diag}[\varepsilon_{n\mathbf{k}}]$, and
	\begin{equation}
\begin{split}
	 \mathcal{G}_\mathbf{k}=&[|g_{+\mathbf{k}}\rangle,|g_{0\mathbf{k}}\rangle,|g_{-\mathbf{k}}\rangle]\,,\\
 \tilde{\mathcal{G}}_\mathbf{k}=&[\langle\tilde{g}_{+\mathbf{k}}|,\langle\tilde{g}_{0\mathbf{k}}|,\langle\tilde{g}_{-\mathbf{k}}|]^T\,.
\end{split}
	\end{equation}
The Hamiltonian (\ref{eq_BdG}) can be diagonalized as $\bar{H}_{\mathbf{k}}=\sum_\mathbf{k}E_\mathbf{k}(\bar{\gamma}_{\mathbf{k}\uparrow}\gamma_{\mathbf{k}\uparrow}+
\bar{\gamma}_{-\mathbf{k}\downarrow}\gamma_{\mathbf{-k}\downarrow})-\sum_\mathbf{k}E_\mathbf{k}$, where $\gamma_{\mathbf{k}\sigma}=[\gamma_{\mathbf{k}\alpha\sigma}]^T$, $E_\mathbf{k}=\rm{diag}[E_{+\mathbf{k}},E_{0\mathbf{k}},E_{-\mathbf{k}}]$, $E_{\pm\mathbf{k}}=\sqrt{\varepsilon_{\mathbf{k}}+\Delta_s^2}\pm\Delta_d$, $E_{0\mathbf{k}}=\Delta_1$, with $\Delta_d\equiv(\Delta_1-\Delta_2)/2$ and $\Delta_s \equiv (\Delta_1+\Delta_2)/2$. The quasiparticle operators are given by
\begin{equation}
\begin{split}
\bar{\gamma}_{\boldsymbol{k} \uparrow}=&u_{\boldsymbol{k}} c_{\boldsymbol{k} \uparrow}^{\dagger}-v_{\boldsymbol{k}} c_{-\boldsymbol{k} \downarrow}\,, \\
\bar{\gamma}_{-\boldsymbol{k} \downarrow}=&v_{\boldsymbol{k}} c_{\boldsymbol{k} \uparrow}+u_{\boldsymbol{k}} c_{-\boldsymbol{k} \downarrow}^{\dagger}\,, \\
\gamma_{\boldsymbol{k} \uparrow}=&u_{\boldsymbol{k}} c_{\boldsymbol{k} \uparrow}-v_{\boldsymbol{k}} c_{-\boldsymbol{k} \downarrow}^{\dagger}\,, \\
\gamma_{\boldsymbol{k} \downarrow}=&v_{\boldsymbol{k}} c_{\boldsymbol{k} \uparrow}^{\dagger}+u_{\boldsymbol{k}} c_{-\boldsymbol{k} \downarrow}\,,
\end{split}\label{eq_qsp}
\end{equation}
where the coefficients $u_\mathbf{k}$ and $v_\mathbf{k}$ take values of
\begin{equation}
\begin{split}
u_{\mathbf{k}}=&\frac{1}{\sqrt{2}}\left[\begin{array}{ccc}
\cos \frac{\phi_{\mathbf{k}}}{2} & 0 & -\cos \frac{\phi_{\mathbf{k}}}{2} \\
0 & 1 & 0 \\
\sin \frac{\phi_{\mathbf{k}}}{2} & 0 & \sin \frac{\phi_{\mathbf{k}}}{2}
\end{array}\right],\\
v_{\mathbf{k}}=&\frac{1}{\sqrt{2}}\left[\begin{array}{ccc}
\sin \frac{\phi_{\mathbf{k}}}{2} & 0 & -\sin \frac{\phi_{\mathbf{k}}}{2} \\
0 & 1 & 0 \\
\cos \frac{\phi_{\mathbf{k}}}{2} & 0 & \cos \frac{\phi_{\mathbf{k}}}{2}
\end{array}\right]\,,
\end{split}\label{eq_bog}
\end{equation}
with $\cos \frac{\phi_{\mathrm{k}}}{2}=\frac{1}{\sqrt{2}} \sqrt{1+\frac{\epsilon_{\mathrm{k}}}{\sqrt{\epsilon_{\mathrm{k}}^{2}+\Delta_{\mathrm{s}}^{2}}}}$, and $\sin \frac{\phi_{\mathrm{k}}}{2}=\frac{1}{\sqrt{2}} \sqrt{1\!-\!\frac{\epsilon_{\mathrm{k}}}{\sqrt{\epsilon_{\mathrm{k}}^{2}+\Delta_{\mathrm{s}}^{2}}}}$. From Eqs. (\ref{eq_qsp}) and (\ref{eq_bog}), the gap equations are obtained;
\begin{equation}
\begin{split}
\Delta_{1}=&\frac{U}{4 N_{\mathrm{c}}} \sum_{\mathbf{k}} [t_{+, \mathbf{k}} \sin \phi_{\mathbf{k}}+t_{-, \mathbf{k}} ]+\frac{U}{4} \tanh \frac{\beta \Delta_{1}}{2}\,,\\
\Delta_{2}=&\frac{U}{2 N_{\mathrm{c}}} \sum_{\mathbf{k}} [t_{+, \mathbf{k}} \sin \phi_{\mathbf{k}}-t_{-, \mathbf{k}} ]\,,
\end{split}\label{eq_order}
\end{equation}
where  $\quad t_{\pm, \mathbf{k}}=\frac{1}{2} (\tanh \frac{\beta E_{+, \mathbf{k}}}{2} \pm \tanh \frac{\beta E_{-, \mathbf{k}}}{2} )$ with the inverse temperature $\beta=1/k_B T$, and $N_c$ is the number of unit cells and we also have $\Delta_1=\Delta_3$ as a result of the sublattice (chiral) symmetry of the Hamiltonian. Note that the flat-band contribution only enters in the gap equations for the order parameter $\Delta_1(\Delta_3)$, due to the destructive interference on sublattice-2 \cite{Julku2016}.

We numerically solve the gap Eq. (\ref{eq_order}). The results are shown in Fig. \ref{fig3}. Here we only focus on the zero temperature solutions as the simplest first approach. Figs. \ref{fig3}(a) and \ref{fig3}(b) show the pairing orders as functions of two-body loss $\Gamma$. Both ${\rm{Re}}\Delta_1$ and ${\rm{Re}}\Delta_2$ are suppressed by the two-body loss, and vanish after crossing a critical point. However, when the two-body loss is large enough, ${\rm{Re}}\Delta_1$ and ${\rm{Re}}\Delta_2$ acquire nontrivial values again and are even enhanced as the loss increases, which can be attributed to the localization-enhanced pairing due to the quantum Zeno effect for strong dissipation. This is reminiscent of the behavior of the pairing field in a single-band BCS system \cite{Yamamoto2019}.

Although the pair field is suppressed by the loss, the gap $\Delta_d$ in the mean-field spectrum is instead enhanced, due to slower decrement occurring in the pair field ${\rm{Re}}\Delta_1$ for sublattice-1 than for ${\rm{Re}}\Delta_2$, as shown in Fig. \ref{fig3} (d). This implies that the superfluidity is more robust against the two-body loss perturbation for the flat bands.

\section{Superfluid weight}\label{sec_sw}

To compute the superfluid weight, we expand the free energy $\mathfrak{F}(\mathbf{A})$ to second order $\mathfrak{F}(\mathbf{A}) \approx \mathfrak{F}_0+\frac{1}{2}V [D_s]_{ij}A_iA_j$, where $V$ is the system area, $D_s$ is the superfluid weight, and $\mathfrak{F}(\mathbf{A})=-\frac{1}{\beta}\ln[e^{-\beta\bar{H}_{\mathbf{k}}}]$, where $\beta=1/k_B T$ is the inverse temperature. Thus the superfluid weight is given by
\begin{equation}
[D_{s}]_{i,j}=\left.\frac{1}{V \hbar^{2}} \frac{\partial^{2} \mathfrak{F}}{\partial q_{\mathrm{i}} \partial q_{\mathrm{j}}}\right|_{\mu, \mathbf{q}=0},
\end{equation}
where $i,j=x,y$ are spatial indices.

For multiband superconducting phase, geometrical contribution enters the superfluid weight in the form of Fubini-Study metric integral of the quantum state manifold, beyond the conventional Landau-Ginzburg formalism. Specifically, For Hamiltonian Eq. (\ref{eq_BdG}), the superfluid weight reads
\begin{widetext}
	\begin{equation}
	 \begin{aligned}
D_s^{i, j}&=\frac{1}{A \hbar^{2}} \sum_{\mathbf{k}}[-2 t_{+, \mathbf{k}} \cos \phi_{\mathbf{k}} \partial_{k_{i}} \partial_{k_{j}} \epsilon_{\mathbf{k}}-\frac{4 t_{-, \mathbf{k}}}{E_{+, \mathbf{k}}-E_{-, \mathbf{k}}} \partial_{k_{i}} \epsilon_{\mathbf{k}} \partial_{k_{j}} \epsilon_{\mathbf{k}}\\
&+2 \Delta_{1} (\tanh \frac{\beta \Delta_{1}}{2}+t_{+, \mathbf{k}} \sin \phi_{\mathbf{k}}+t_{-, \mathbf{k}} ) ( \langle\partial_{k_{i}} \tilde{s}_{\mathbf{k}} | \partial_{k_{j}} s_{\mathbf{k}} \rangle+ \langle\partial_{k_{j}} \tilde{s}_{\mathbf{k}} | \partial_{k_{i}} s_{\mathbf{k}} \rangle ) \\
& -\Delta_{1}^{2} \langle\partial_{k_{i}} \tilde{s}_{\mathbf{k}} | s_{\mathbf{k}} \rangle \langle \tilde{s}_{\mathbf{k}} | \partial_{k_{j}} s_{\mathbf{k}} \rangle f(\mathbf{k})-\Delta_{1}^{2} ( \langle\partial_{k_{i}} \tilde{s}_{\mathbf{k}} | c_{\mathbf{k}} \rangle \langle \tilde{c}_{\mathbf{k}} | \partial_{k_{j}} s_{\mathbf{k}} \rangle+(i \leftrightarrow j) ) g(\mathbf{k}) ].
\end{aligned}
	\end{equation}
\end{widetext}
The flat band has contribution,
\begin{equation}
\left. [D_{s} ]_{i, j}\right|_{\mathrm{f.b}}=\frac{\Delta_1}{\pi \hbar^2}\tanh\frac{\beta\Delta}{2}\left.\mathcal{B}_{ij}\right|_{\mathrm{f.b}}\,,
\label{eq_fbsw}
\end{equation}
where $\mathcal{B}_{ij}=(2\pi)^{-1}\int_{\rm{B.Z.}}d^2\mathbf{k} \mathfrak{G}_{ij}$ is the Brillouin-zone integral of the tensor, with its real part the quantum metric. The quantum metric is defined as $\mathcal{M}_{ij}={\mathrm{Re}}~\mathfrak{G}_{ij}$, where
\begin{equation}
\begin{split}
\mathfrak{G}_{ij}=&\frac{1}{2}[\langle\partial_{k_{i}} \tilde{g}_{0 \mathbf{k}} | \partial_{k_{j}} g_{0 \mathbf{k}} \rangle+\langle\partial_{k_{j}} \tilde{g}_{0 \mathbf{k}} | \partial_{k_{i}} g_{0 \mathbf{k}} \rangle\\
-&\langle\partial_{k_{i}} \tilde{g}_{0 \mathbf{k}}|g_{0 \mathbf{k}} \rangle \langle \tilde{g}_{0 \mathbf{k}}| \partial_{k_{j}} g_{0 \mathbf{k}} \rangle \!-\! \langle\partial_{k_{j}} \tilde{g}_{0 \mathbf{k}}|g_{0 \mathbf{k}} \rangle \langle \tilde{g}_{0 \mathbf{k}}| \partial_{k_{i}} g_{0 \mathbf{k}} \rangle]\,.
\end{split}
\end{equation}
We note that Eq. (\ref{eq_fbsw}) takes similar form with its Hermitian counterpart \cite{KSun2011}, but with the generalized non-Hermitian metric (see Appendix \ref{appa}) and in a biorthogonal formalism.

\begin{figure}[htbp]
	\centering
	\includegraphics[width=\textwidth]{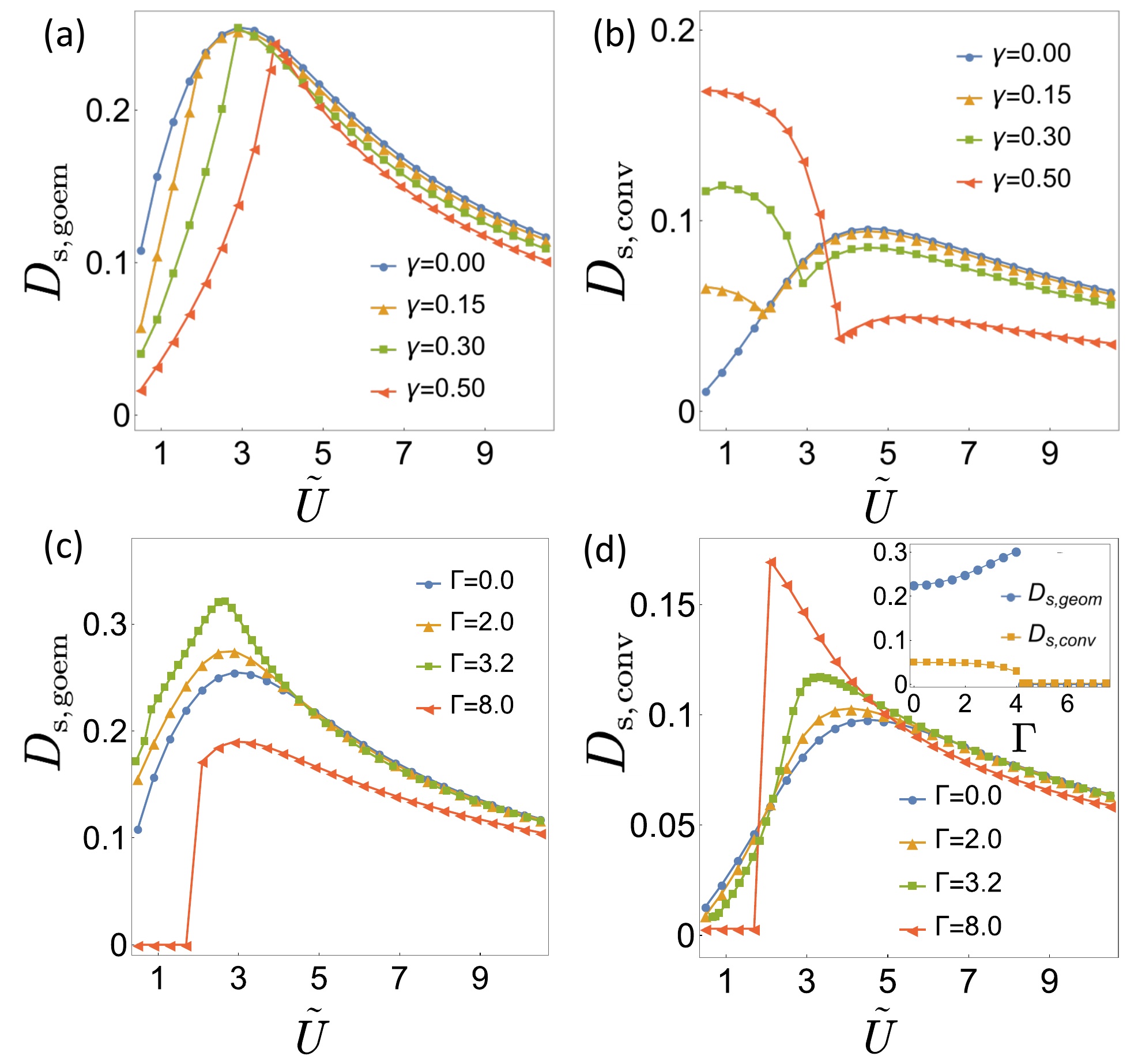}
	\caption{ (a) Geometric and (b) conventional contribution in the superfluid weight at zero temperature with only non-reciprocity. (c) and (d) The geometric and conventional contribution in the superfluid weight at zero temperature with only two-body loss. The inset in (d) shows the superfluid weights as functions of two-body loss strength $\Gamma$ at $\tilde{U}=1.8$.}
	\label{fig4}
\end{figure}

In Fig. \ref{fig4}, we plot the geometric contribution and the conventional contribution, respectively. Here we only focus on the diagonal components of the superfluid weight tensor $[D_s]_{xx}=[D_s]_{yy}\approx D_s$, since the off-diagonal components $[D_s]_{xy/yx}$ are small. In the absence of non-Hermiticity, the superfluid weight $D_s$  is enhanced as $\tilde{U}$ increases in the weak-coupling regime. But $D_s$ decreases after peaking at $\tilde{U}\sim 4t$ with $\tilde{U}$ further increases. The non-monotonicity of $D_s$ can be understood from the viewpoint of pseudopotential theory and a crossover from unbound Cooper pairs to preformed fermions (or Bose liquid). Here we give a rough estimation. The s-wave scattering length $a_s$ is rebuilt from the relation $\tilde{U}=4\pi\hbar^2a_s/M$ with $M$ the mass of atoms. Taking $\hbar=M_r=1$, and the Fermi momentum $k_F\sim \pi$, we have $1/k_Fa_s\sim -1$ for $-\tilde{U}=-4$. After reaching the unitarity regime ($-1<1/k_Fa_s<1$), the system will experience a crossover from the unbound Cooper pairs to the bounded pairs (molecules of fermions) with small pair size. The pairing happens on site, disordering the phase coherence of the pair field, thus leading to a decrease of $D_s$. In the strong coupling regime, the superfluid weight is more robust to the non-Hermitian perturbations, as shown in Fig. \ref{fig4}.

Notably, we find that the nonreciprocity enhances the superfluid weight in the weak-coupling regime, as shown in Fig. \ref{fig4} (b) for the nonreciprocal BdG Hamiltonian.
Here we give an intuitive insight on this enhancement. As revealed in Sec. \ref{sec_free}, in the presence of nonreciprocity, the system has larger density of states near the Fermi surface. Furthermore, the nonreciprocity also enhances the flat band localization. As the Cooper pairs form around the Fermi surface, the superconductivity is more easily induced thus the coherence of the superconducting phase is promoted for the conventional part. In contrast, the geometric part decreases as nonreciprocity increases in the weak-coupling regime. On the other hand, for the Hamiltonian only with two-body loss, the superfluid weight is decreased by the loss, as shown in Fig. \ref{fig4} (d).

Furthermore, we also find that the overall behavior of the geometric part mostly depends on the gap parameter $\Delta_d$ after comparing the results in Fig. \ref{fig4} and $\Delta_d$ under same system parameters.

\section{Discussion and conclusion}\label{sec_conclusion}
The localization properties discussed in Sec. \ref{sec_free} can be probed in the experiment. By preparing the system on the flat band, the real space distribution could be directly revealed by the single-atom-resolved fluorescence imaging \cite{Bloch2012,Taie2015}. Alternatively, as propose in Ref. \cite{Ozawa2019}, the position variance could be extracted by the excitation rate into other levels after a driving $H_D=2E \hat x\cos(\omega t)$ is applied, $\Gamma=\frac{2\pi E^2}{\hbar^2} \mathrm{Var}(\hat x)$. Furthermore, the flat band localization can also be reflected in the dynamical properties \cite{ZHZhang2016,LJin2019}. 

In the experiment, the tunneling strength could be tuned by the lattice depth. Here we estimate the tunneling strength $t \sim 2\pi\times100~\rm{Hz}$. If the primary lattice is coupled with the auxiliary lattice with strength $\Omega\sim 2\pi\times 1~\rm{kHz}$, a non-reciprocity $\delta=0.5 t$ can be generated with decay rate $\kappa \sim 2.5\Omega$ in auxiliary lattice (The atom decay could be engineered with spin-dependent dissipation in clod atomic systems in a large parameter regime \cite{JLi2019}). In two dimensions, the superfluid weight is related with the Berezinskii-Kosterlitz-Thouless transition. A larger superfluid weight  at zero temperature usually implies a higher transition temperature \cite{FXie2020}. Thus the superfluid weight can be estimated by the critical temperature in the experiment. The critical temperature is estimated  to be $k_B T_c \sim 0.1 t$ for $U=4t$ in the Hermitian limit \cite{Julku2016}.

In conclusion, we have proposed a cold atomic setup with uniform s-wave interaction and atom loss, which is captured by a non-Hermitian Hamiltonian. We started by elucidating unique localization properties of the Bloch bands as an interplay of skin effect and the flat-band destructive interference. Then with the inclusion of complex s-wave interaction, we revealed criticality of the non-Hermitian superfluidity related with the flat band within the framework of a mean-field BCS theory, in terms of both order parameters and the superfluid weight. We built a relation between the superfluid weight and the geometry of the Bloch states manifold in the non-Hermitian case. And we have also showed that the non-reciprocity would optimize the superfluid weight. Our work provides an example that the non-reciprocity in non-Hermitian systems can bring important physical implications.

\acknowledgments
We thank Dan-Bo Zhang and Yu-Guo Liu for useful discussions. This work was supported by the Key-Area Research and Development Program of GuangDong Province (Grant No. 2019B030330001), the National Natural Science Foundation of China (Grants No. 12074180 and No. U1801661), and the Key Project of Science and Technology of Guangzhou (Grant No. 201804020055).

\begin{appendix}
\section{Quantum metric tensor}\label{appa}

In this section, we introduce the non-Hermitian quantum metric tensor, following the definition in Ref. \cite{DJZhang2019}. For the set of the quantum states $\{|g_{n\mathbf{k}}\rangle\}$ on the Brillouin
zone(BZ) torus, the density matrix is defined as,
\begin{equation}
\rho_n(\mathbf{k})\equiv|\tilde{g}_{n\mathbf{k}}\rangle\langle g_{n\mathbf{k}}|\,.
\end{equation}
The fidelity between $\rho_n(\mathbf{k})$ and $\rho_n(\mathbf{k}+\delta\mathbf{k})$ is given by
\begin{equation}
F (\rho_{n}(\mathbf{k}), \rho_{n}(\mathbf{k}+\delta \mathbf{k}) )=\operatorname{tr} \sqrt{ |\rho_{n}^{1 / 2}(\mathbf{k}) \rho_{n}(\mathbf{k}+\delta \mathbf{k}) \rho_{n}^{1 / 2}(\mathbf{k}) |}\,.
\end{equation}
The Fubini-Study metric defines a distance of nearby states on the BZ torus,
\begin{equation}
d s^{2}:=2 [1-F (\rho_{n}(\mathbf{k}), \rho_{n}(\mathbf{k}+\delta \mathbf{k}) ) ]\,.
\end{equation}
By expanding the states $|g_{n\mathbf{k}+\delta\mathbf{k}}(\tilde{g}_{n\mathbf{k}+\delta\mathbf{k}})\rangle$ to second order, we have
\begin{equation}
\begin{split}
d s^{2}=&\frac{1}{2} \operatorname{Re} ( \langle\partial_{\mu} \tilde{g}_{n\mathbf{k}}  | \partial_{\nu} g_{n\mathbf{k}}  \rangle+ \langle\partial_{\nu} \tilde{g}_{n\mathbf{k}}|\partial_{\mu} g_{n\mathbf{k}}  \rangle\\
-&2 \langle\partial_{\mu} \tilde{g}_{n\mathbf{k}}|g_{n\mathbf{k}}  \rangle \langle \tilde{g}_{n\mathbf{k}}|\partial_{\nu} g_{n\mathbf{k}}  \rangle ) d k^{\mu} d k^{\nu}\,.\end{split}
\end{equation}
Here to shorten notations we define $\partial_\mu \equiv \partial_{k_\mu}$.
After noting that $\langle\partial_{\mu} \tilde{g}_{n\mathbf{k}}| g_{n\mathbf{k}}  \rangle \langle \tilde{g}_{n\mathbf{k}}  | \partial_{\nu} g_{n\mathbf{k}}  \rangle  =\langle\partial_{\nu} \tilde{g}_{n\mathbf{k}}|g_{n\mathbf{k}}  \rangle \langle \tilde{g}_{n\mathbf{k}}  | \partial_{\mu} g_{n\mathbf{k}}  \rangle $, we can rewrite the quantum metric tensor as
\begin{equation}
\begin{split}
\mathcal{M}_{ij}=&\frac{1}{2}{\mathrm{Re}}[\langle\partial_{k_{i}} \tilde{g}_{n \mathbf{k}} | \partial_{k_{j}} g_{n \mathbf{k}} \rangle+\langle\partial_{k_{j}} \tilde{g}_{n \mathbf{k}} | \partial_{k_{i}} g_{n \mathbf{k}} \rangle\\
-&\langle\partial_{k_{i}} \tilde{g}_{n \mathbf{k}}|g_{n \mathbf{k}} \rangle \langle \tilde{g}_{n \mathbf{k}}| \partial_{k_{j}} g_{n \mathbf{k}} \rangle \!-\! \langle\partial_{k_{j}} \tilde{g}_{n \mathbf{k}}|g_{n \mathbf{k}} \rangle \langle \tilde{g}_{n \mathbf{k}}| \partial_{k_{i}} g_{n \mathbf{k}} \rangle]\,.
\end{split}
\end{equation}

\section{Derivation details of the superfluid weight}
By virtue of the biorthogonal representation, we derive the superfuild weight of the non-Hermitian superfluidity, following the method originally developed in Refs. \cite{Peotta2015,Julku2016,Liang2017}. The first derivative of the free energy $\Omega(\mathbf{q})$ is the current density,
\begin{equation}
\mathbf{J}(\mathbf{q})=\frac{-1}{2 V \hbar} \sum_{\mathbf{k}} \operatorname{Tr}\left[\operatorname{sign}\left(E_{\mathbf{k}}(\mathbf{q})\right) \mathcal{W}_{\mathbf{k}}^{T}(\mathbf{q}) \partial_{\mathbf{q}} H_{\mathbf{k}}(\mathbf{q}) \mathcal{W}_{\mathbf{k}}(\mathbf{q})\right]\,,
\end{equation}
\begin{equation}
-\partial_{\mathbf{q}} H_{\mathbf{k}}(\mathbf{q})=\left[\begin{array}{cc}
\partial_{\mathbf{k}} \varepsilon_{\mathbf{k}-\mathbf{q}} & \partial_{\mathbf{q}} \mathcal{D}_{\mathbf{k}}(\mathbf{q}) \\
-\partial_{\mathbf{q}} \mathcal{D}_{\mathbf{k}}(-\mathbf{q}) & \partial_{\mathbf{k}} \varepsilon_{\mathbf{k}+\mathbf{q}}
\end{array}\right]\,,
\end{equation}
Here we define $\mathcal{D}_\mathbf{k}\equiv \tilde{\mathcal{G}}_{\mathbf{k}-\mathbf{q}} \Delta \mathcal{G}_{\mathbf{k}+\mathbf{q}}$, and
\begin{equation}
\mathcal{W}_{\mathbf{k}}\equiv\left[\begin{array}{cc}
u_\mathbf{k} &-v_{\mathbf{k}}\\v_{\mathbf{k}}&u_{\mathbf{k}}
\end{array}\right].
\end{equation}
Straightforward calculation shows that,
\begin{widetext}
	\begin{equation}
	 \begin{aligned}
\tilde{\mathcal{G}}_{\mathbf{k}_{1}} \Delta \mathcal{G}_{\mathbf{k}_{2}}=\frac{\Delta_{A}}{2}\left[\begin{array}{ccc}
\langle \tilde{s}_{\mathbf{k}_{1}} \mid s_{\mathbf{k}_{2}}\rangle & \sqrt{2}\langle \tilde{s}_{\mathbf{k}_{1}}|c_{\mathbf{k}_{2}}\rangle & \langle \tilde{s}_{\mathbf{k}_{1}} | s_{\mathbf{k}_{2}}\rangle \\
\sqrt{2}\langle \tilde{c}_{\mathbf{k}_{1}}|s_{\mathbf{k}_{2}}\rangle & 2\langle  \tilde{c}_{\mathbf{k}_{1}}|c_{\mathbf{k}_{2}}\rangle & \sqrt{2}\langle  \tilde{c}_{\mathbf{k}_{1}} |s_{\mathbf{k}_{2}}\rangle \\
\langle \tilde{s}_{\mathbf{k}_{1} } | s_{\mathbf{k}_{2}} \rangle & \sqrt{2} \langle \tilde{s}_{\mathbf{k}_{1}}| c_{\mathbf{k}_{2}} \rangle &  \langle \tilde{s}_{\mathbf{k}_{1}} | s_{\mathbf{k}_{2}}\rangle
\end{array}\right]+\frac{\Delta_{B}}{2}\left[\begin{array}{ccc}
1 & 0 & -1 \\
0 & 0 & 0 \\
-1 & 0 & 1
\end{array}\right].
\end{aligned}
	\end{equation}
\end{widetext}
Here we denote $\mathbf{k}_1=\mathbf{k}-\mathbf{q}$ and $\mathbf{k}_2=\mathbf{k}+\mathbf{q}$, and we introduce a two-component spinor $|s_\mathbf{k}\rangle$ (together with $\langle\tilde{s}_\mathbf{k}|$) and its partner $|c_\mathbf{k}\rangle=i\sigma_y|s_\mathbf{k}\rangle$ (together with $\langle\tilde{c}_\mathbf{k}|$),
	\begin{equation}
	|s_\mathbf{k}\rangle=\frac{1}{\sqrt{a_{\mathbf{k}}^2+b_{\mathbf{k}}^2}}\left[\begin{array}{c}
a_\mathbf{k}\\b_{\mathbf{k}}
\end{array}\right],\quad |c_\mathbf{k}\rangle=\frac{1}{\sqrt{a_{\mathbf{k}}^2+b_{\mathbf{k}}^2}}\left[\begin{array}{c}
b_\mathbf{k}\\-a_{\mathbf{k}}
\end{array}\right],
	\end{equation}
	\begin{equation}
	\langle \tilde{s}_\mathbf{k}|=[a_\mathbf{k},b_{\mathbf{k}}]/\sqrt{a_{\mathbf{k}}^2+b_{\mathbf{k}}^2}\,,\quad \langle \tilde{c}_\mathbf{k}|=[b_\mathbf{k},-a_{\mathbf{k}}]/\sqrt{a_{\mathbf{k}}^2+b_{\mathbf{k}}^2}\,.
	\end{equation}
Note that $\langle \tilde{s}_\mathbf{k}(\tilde{c}_\mathbf{k})|$ is not the Hermitian conjugate of $|s_\mathbf{k}(c_\mathbf{k})\rangle$. The derivate of the current density gives rise to the superfluid weight after setting $\mathbf{q}=0$,
\begin{widetext}
\begin{equation}
 [D_{\mathrm{s}} ]_{ij}=\left.\frac{1}{V \hbar^{2}} \frac{\partial^{2} \mathfrak{F}}{\partial q_{i} \partial q_{j}}\right|_{\mathbf{q}=0}=\left[D_{\mathbf{s},\text{conv }}\right]_{i, j}+\left[D_{\mathbf{s},\text{geom }}\right]_{i, j}\,,
\end{equation}
\begin{equation}
[D_{\mathrm{s}, \mathrm{conv}}]_{ij}=\frac{2}{V \hbar^{2}} \sum_{\mathbf{k}} \operatorname{Tr} [ (v_{\mathbf{k}} \frac{1}{e^{-\beta E_{\mathbf{k}}+1}} v_{\mathbf{k}}^{T}+u_{\mathbf{k}} \frac{1}{e^{\beta E_{\mathbf{k}}+1}} u_{\mathbf{k}}^{T} ) \partial_{k_{i}} \partial_{k_{j}} \varepsilon_{\mathbf{k}} ]\,,
\end{equation}
\begin{equation}
\begin{split}
[D_{\mathrm{s}, \mathrm{geom}}]_{ij}
=\frac{1}{V \hbar^{2}} \{ 2 \sum_{\mathbf{k}} \operatorname{Tr} & [ (u_{\mathbf{k}} v_{\mathbf{k}}^{T}-u_{\mathbf{k}} \frac{1}{e^{\beta E_{\mathbf{k}} +1}} v_{\mathbf{k}}^{T}-v_{\mathbf{k}} \frac{1}{e^{\beta E_{\mathbf{k}}+1}} v_{\mathbf{k}}^{T} ) \partial_{q_{i}} \partial_{q_{j}} \mathcal{D}_{\mathbf{k}}(\mathbf{q}=0) ] \\
& -\frac{1}{2} \sum_{\mathbf{k}} \sum_{a, b} [T_{\mathbf{k}} ]_{a, b} [N_{\mathbf{k}, i} ]_{a, b} [N_{\mathbf{k}, j} ]_{b, a} \}\,,
\end{split}
\end{equation}
\end{widetext}
where $N_{\mathbf{k}, i}=\mathcal{W}_{\mathbf{k}}^{T}(\mathbf{q}=0) \partial_{q_{i}} \bar{H}_{\mathbf{k}}(\mathbf{q}=0) \mathcal{W}_{\mathbf{k}}(\mathbf{q}=0)$, and  the off-diagonal components of $T_{\mathbf{k}}$ $[T_{\mathbf{k}} ]_{a, b}= (\tanh (\beta E_{\mathbf{k}} / 2 ) ]_{a, a}- [\tanh(\beta E_{\mathbf{k}}/2 ) ]_{b, b})/ ([E_{\mathbf{k}} ]_{a, a}- [E_{\mathbf{k}} ]_{b, b})$, and the diagonal components $[T_{\mathbf{k}} ]_{a, a}= [ \beta/2 \cosh ^{2} (\beta E_{\mathbf{k}}/2 )]_{a, a}$.

\end{appendix}

\end{document}